\documentclass[12pt,preprint]{aastex}
\begin{document}

\title{The Fornax dwarf galaxy as a remnant of recent dwarf-dwarf merging
in the Local Group}

\author{C. Yozin and K. Bekki} 
\affil{
ICRAR,
M468,
The University of Western Australia
35 Stirling Highway, Crawley
Western Australia, 6009, Australia
}

\begin{abstract}

We present results from the first numerical analysis to support the hypothesis, first proposed in Coleman et al. (2004),  
that the Fornax dwarf galaxy was formed from the minor merging of two dwarfs
about 2 Gyr ago. Using orbits for the Fornax dwarf that are consistent with the latest proper motion measurements, 
our dynamical evolution models show that the observed asymmetric shell-like substructures
can be formed from the remnant of a smaller dwarf during minor merging. These models also predict the formation of
diffuse stellar streams.  We discuss how
these stellar substructures depend on model parameters of dwarf-dwarf merging, 
and how the intermediate-age subpopulations found in the vicinity of these substructures may be formed from gas accretion
in the past merger events. We also suggest that one of Fornax's GCs originates from a
merged dwarf companion, and demonstrate where as yet undetected tidal streams
or HI gas formed from the dwarf merging may be found in the outer halo of the Galaxy.
\end{abstract}


\section{Introduction}
The Fornax galaxy possesses many unusual characteristics that have warranted far greater investigation than might be expected for a typical dwarf spheroidal (dSph). Lying in the Local Group at a distance of 138kpc (Saviane et al. 2000) and receding at 53.3kms$^{-1}$ (Walker et al. 2007), the foremost distinguishing features are six Globular Clusters (GC), resulting in one of the highest known specific frequencies of all galaxies (29, from van den Bergh 1998). Although these GCs are associated with Fornax (Mateo et al. 1991), with distances to the galaxy center ranging between 0.24 and 1.60kpc, there exists an atypical spread in metallicity and age (Buonanno et al. 1999, and Strader et al. 2003). Moreover, dynamical friction acting on the GCs in the dark matter (DM) dominated Fornax should have resulted in their assimilation into the Fornax nucleus since their formation more than $\sim$10 Gyr ago. Tidal interaction with the Milky Way (MW) has been invoked to explain the latter by Oh et al. (2000); further studies by Goerdt et al. (2006) propose that a cored DM halo would increase the effective sinking time for GCs, while Cowsik et al. (2009) asserts dynamical friction to be weaker than previously thought.
\par
The above discussion lies in parallel but separate from recent photometric analyses by Coleman et al. (2004, 2005). These studies have interpreted statistically-significant (4.5$\sigma$) stellar overdensities as evidence of an extant shell substructure, with the most prominent shell 'sheets' aligned with the major axis of Fornax and extending beyond the tidal radius ($r_t = 71\pm4$ arcmin, Irwin \& Hatzidimitriou 1995) of Fornax . Tidal interaction with the MW is discarded as a causative factor, since the predominantly intermediate (2 Gyr) age of the overdensity subpopulation contrasts with the otherwise mixed aged general population of Fornax (Buonanno et al. 1999, Stetson et al. 1998). Quinn (1984) showed that low-energy minor mergers can result in shell-like substructures, and a gas-rich companion galaxy may also provide the fuel source for a past starburst, given the present lack of $H_I$ gas observed in Fornax. Olszewski et al. (2006) argue however that the metallicity in the overdensities bear the signature of Fornax itself.
\par
In both debates, the influence of the Milky Way (MW) potential has been contemplated with respect to the assumed orbit of Fornax. Successive measurements for the present day proper motion of Fornax have been conducted recently by Dinescu et al. (2004), Piatek et al. (2007), Walker et al. (2008), and M\'endez et al. (2011), hereafter D04, P07, W08 and M11 respectively. 
M11 gives a high-energy orbit with an orbital period $\sim$21 Gyr, while D04, P07 and W08 imply a low-energy orbit with orbital periods ranging between 5-8 Gyr. In all cases, Fornax is conceived to have been at perigalacticon in the last $\sim$100-300 Gyr. M\'endez et al. (2011) propose that this triggered an enhanced star formation at this period, and perhaps a similarly induced starburst $>$10 Gyr ago, thus supporting their hyper-extended orbit (if the merger scenario can explain the remaining intermediate subpopulation). By contrast, the minimum galactocentric radial distance of 131-148kpc leads Coleman et al. (2005) to assert the relative insignificance of tidal forces at this separation. To date, no numerical studies have been conducted to assert the interaction and merger history of Fornax, using the latest measurements of its proper motion.
\par
The purpose of this {\it LETTER} is to test the hypothesis of Coleman et al. (2004) that Fornax is a
remnant of recent dwarf merging.
We show that the overdensities are evidence of a shell substructure
formed by the most recent merger event.
We also discuss how the recent dwarf merging is important for (i) the formation
of intermediate-age ($\sim$2 Gyr) stars in the core of Fornax
and (ii) physical properties of its unusual globular cluster system. 
We make predictions for future observations 
that would constrain the present range of proper motions and further support the merger hypothesis.

\section{The model}

Our investigation assumes that the present day Fornax dSph is the product of the merger of two dwarfs, referred to hereafter as 
Fornax A (host) and B (companion) respectively.
Our merger model is developed in two stages:  
first, we evaluate the orbital history of Fornax for a given Galactic
potential and a mass model for Fornax A/B (\S2.1) by using the 
"backward integration scheme'' used in similar simulations
(e.g., Diaz \& Bekki 2012, hereafter DB12). In the second step (\S2.2) we introduce an collisionless N-body representation of Fornax B, and investigate its dynamical evolution under the gravitational influence of the Galaxy and Fornax A.

\subsection{Orbit Integration} 
We adopt the same Galaxy model as used in our previous simulations on the evolution
of the Magellanic Cloud (DB12). The dominant component is the extended halo, for which we adopt a NFW density distribution (Navarro et al. 1996), up to a virial radius $R_{vir}$ of 175kpc:
\[
	\rho_{NFW}(r) = \frac{\rho_0}{(cr/R_{vir})(1+cr/R_{vir})^2},
\]
from which the virial mass $M_{vir}$ is derived as:
\[
		M_{vir} = 4\pi\rho_0(R_{vir}/c^3)(ln(1+c)-c/(1+c)) = 1.30\times10^{12}M_{\odot},
\]
where $r$ is the spherical radius, $\rho_0$ is the characteristic density and $c$ is a concentration parameter ($=12$). The disk uses the potential of Miyamoto \& Nagai (1975), which has the form:
\[
	\Phi_{disk} = -\frac{GM_d}{\sqrt{R_2 + (a + \sqrt{z_2 + b_2})^2}},
\]
where the total mass of the disk $M_d=5.0\times10^{10}M_{\odot}$ (Binney \& Tremaine 2008), $R=\sqrt{x^2 + y^2}$, and $a$ and $b$ are scale parameters with values 3.5kpc and 0.35kpc respectively. Finally, the Galactic bulge uses the spherical potential model of Hernquist (1990):
\[
	\Phi_{bulge} = -\frac{GM_b}{r+c_b}
\]
where the mass of the bulge $M_b$ is $0.5\times10^{12}M_{\odot}$ and $c_b$ = 0.7kpc (Binney \& Tremaine 2008). The total mass of the MW, up to a limit radius of 300kpc, is thus calculated as $\sim1.73\times10^{12}M_{\odot}$. 
Fornax A is assumed to have a Plummer potential:
\[
	\Phi_{FA}(r) = -\frac{GM_{FA}}{\sqrt{r_2+{a_{FA}}^2}},
\]
where $M_{FA}$ is the total mass of ($10^9 M_{\odot}$, consistent with the latest observations
(e.g., de Boer et al. 2011), and $a_{FA}$ is a softening radius of 0.6kpc. We consider the spherical Plummer model sufficient for the slightly 
oblate extant Fornax dSph; Hernquist \& Quinn (1989) show a complex 
dependence of the shell structure alignment on primary galaxy potential, 
but we consider this dependence sufficiently small to ignore
within the scope of this preliminary study. 
\par
The time integration of the equation of motion
is performed by using 2nd-order
leap-frog method (Murai \& Fujimoto 1980) with a time step interval of $\sim 1.4 \times 10^6$ yr. For initial space velocities, we used the proper motions of P07 and M11 as representative of the low- and high-energy orbits respectively. The calculated error in the measurements of P07 were sufficiently low to justify the use of space velocities (in the galactocentric rest frame) derived from the nominal proper motion. The measurements of M11 by contrast are accompanied by significant confidence limits, in which case we separately considered the nominal (M11), lowest- (M11L) and highest- (M11H) energy combinations from this data to attain the corresponding space velocities ($V_X$, $V_Y$, $V_Z$) in Table 1. We use the coordinate system adopted in DB12, in which Fornax is currently located at position $(X,Y,Z) = (-39.0, -48.0, -126.0)$ kpc, as derived from van den Bergh (1999). For each orbit, the integration scheme is utilised until the time ($T_M$) at which a merger is judged to have commenced. To establish the time varying dependence of our postulated mergers, and accommodate the age range of young stars in Fornax (e.g. Fig.4, Tsujimoto 2011), we vary the merger time among models from $T_M$ = -3.5 to -2 Gyr, where $T = 0$ is the present time.

\subsection{N-Body model and merger parameters}
Fornax B is assumed to be a dwarf disk
galaxy embedded in a massive dark matter halo, and is therefore represented here by N-body particles in
a self-consistent manner. We adopt the dwarf disk
model adopted in our previous simulations on dynamical evolution
of dwarf disk galaxies around the Galaxy, the details of which are given in Bekki \& Yong (2012).

The total mass of Fornax B ($M_{\rm t,dw}$)
is assumed as 5\% of Fornax A (i.e., $M_{\rm t,dw}= 5 \times 10^7 M_{\odot}$). The observed average metallicity of Fornax is consistent with the merger of two satellites (Fornax A and B) in a 20:1 stellar mass ratio assuming that the two merging satellites conform to the luminosity-metallicity relationship of MW dSph satellites (e.g., Kirby et al. 2011).
The density profile of the dark matter halo
with a total mass of $M_{\rm dm, dw}$ is represented by that proposed by
Salucci \& Burkert (2000):
\[
	{\rho}_{\rm dm}(r)=\frac{\rho_{\rm dm,0}}{(r+a_{\rm dm})(r^2+{a_{\rm dm}}^2)},
\]
where $\rho_{\rm dm,0}$ and $a_{\rm dm}$ are the central dark matter
density and the core (scale) radius, respectively.
The mass-ratio of $M_{\rm dm, dw}$ to $M_{\rm t,dw}$ is set to be 0.9, 
while $a_{\rm dm}$ is varied for different models.
The stellar component of the dwarf is modelled as a bulge-less stellar disk
with the total mass of $M_{\rm s,dw}$ and the size of $R_{\rm s, dw}$.
The radial ($R$) and vertical ($Z$) density profiles of the stellar disk are
assumed to be proportional to $\exp (-R/R_{0}) $ with scale
length $R_{0} = 0.2R_{\rm s, dw}$ and to ${\rm sech}^2 (Z/Z_{0})$ with scale
length $Z_{0} = 0.04R_{\rm s, dw}$, respectively.
The gravitational softening length  for the dark
matter and stellar disk is set to be 2.7pc.
\par
In addition to the
rotational velocity caused by the gravitational field of disk
and dark halo components, the initial radial and azimuthal
velocity dispersions are assigned to the disc component according to
the epicyclic theory with Toomre's parameter $Q$ = 1.5. 
The mass-ratio of $M_{\rm s, dw}$ to $M_{\rm t,dw}$ is set to be 0.1.
The spin of the  dwarf galaxy
is specified by two angles $\theta$ and
$\phi$ (in units of degrees).
Here, $\theta$ is the angle between
the $z$-axis and the vector of the angular momentum of the disk,
and $\phi$ is the azimuthal angle measured from the $x$-axis to
the projection of the angular momentum vector of the disk onto
the $x$-$y$ plane. The initial position of Fornax B, with respect to Fornax A, is described with:
\[
	{\bf X}_{AB} = (r_{AB}\cos(\Phi),0,r_{AB}\sin(\Phi)),
\]
where $r_{AB}$ is the initial radial separation and $\Phi$ is an inclination angle; similarly, the velocity is:
\[
	{\bf V}_{AB} = (f_rV_c,f_tV_c,0),
\]
where $V_c$ is the circular velocity at $r_{AB}$, and $f_r$ and $f_t$ are radial and tangential component factors respectively. The locations of the overdensities from Coleman et al. (2004, 2005) would appear to describe an aligned shell, which is most often imposed by radial or near-radial collisions (Hernquist \& Quinn 1989). The relative sparseness of the Local Group, and Fornax's location on the periphery of those satellites considered bound to the Galaxy, would result in galaxy interactions tending towards those observed in field rather than cluster environments. We thus constrained the trajectory of Fornax B with respect to A to a low-energy, infalling parabolic orbit, with $f_r = -0.5$ and $f_t = 0.1$.  The morphology of a shell substructure is also sensitive to the initial orientation of the encounter: the sharp-edged features of the shell are phase-wrapped sheets as seen edge-on, with their overall distribution (the symmetry plane) seen face-on. We assumed that the overdensities fit this perspective, and thus we determined $\theta=135$, $\phi=0$ and $\Phi = 30^{\circ}$, which together promote a face-on view from a heliocentric location at $T = 0$. 
\par
The dynamical evolution of Fornax B was investigated for a wide extent of the contingent parameter space described by the variables here defined, with models consisting of 200,000 particles (split equally between the stellar and DM components). For parameter sets where the stellar distribution is comparable with extant observations, we confirm the results with higher resolution 1,000,000 particle models. All simulations are implemented with an original code (e.g., Bekki 2011), which executes with the parallel computing architecture CUDA, on a PC cluster installed with Nvidia GTX580 GPUs at the University of Western Australia. 

\section{Results}
Figure 1 shows the backwards integrated orbits of Fornax A for the nominal proper motions of P07 and M11, and conveys the wide discrepancy in their respective proximity to the MW. For a merger instantiated at ($T_M = -3.5$) using the proper motion of P07 we find, in the dynamic stellar distribution, overdensities are developed for a specific parameter subset of Fornax B that we define as Model 1 (Table 1). The top row of Figure 2 gives the projected stellar distribution of Model 1, with a Gaussian-convolved image of Fornax from Coleman et al. (2005) for comparison; the outer overdensity is highlighted together with the inferred axis of symmetry. At $T = -1.4$, a classical type 1 (Prieur 1990 classification) 'bow-tie' structure is prominent. The regions of maximal stellar density are further elucidated by first dividing the on-sky distribution projection into quadrants, and then counting the number of stars that lie within each equal width bin of radius from Fornax A.  Accordingly, the top-left panel of Figure 3 shows salient maximal densities at $1^{\circ}$ to $1.5^{\circ}$ arranged in opposing quadrants, analogous to Coleman et al. (2005). As in other studies (e.g. Weil et al 1996), these peaks exist briefly ($\sim$0.1 Gyr) before diminishing through mixing. Fornax B is not completely disrupted, but we find it to first reside at the maximal stellar density region, before later descending to Fornax A's stellar core. Notably, we find a {\it bimodality} in the heliocentric velocity of the Fornax B stellar remnant up to a radius of $\sim0.7^{\circ}$; this can be compared with a similar bimodality observed for the metal-poor stars of the Fornax dSph (Battaglia et al. 2006), as conveyed in the right-side panels of Fig. 3).
\par
When an otherwise identical model is run but with $T_M = 2.1$ Gyr (Model 2), the anticipated overdensities are in fact {\it not} reproduced at $T =  0$. Tidal influence on Fornax is deemed strongest therefore during its approach to perigalacticon, and to recover overdensities at $T = 0$ for this $T_M$, Fornax B is required to be more compact in its original state. With a decreased disk $R_{\rm s, dw}$ and halo core radius $a_{\rm dm}$, Model 3 is conveyed in Fig. 2 and Fig. 3 to possess a distinct overdense remnant (the surviving stellar core of Fornax B) at $T = 0$, which is again comparable qualitatively with Fornax. Taking a broader view of the stellar distribution, the left-side panels of Figure 4 show a tidal stream in Models 1 and 3, which extends up to a length of $\sim30^{\circ}$ in alignment with the prior orbit of Fornax, and for a further $\sim20^{\circ}$ in the opposing direction. 
\par
In contrast, our models with high energy orbits (M11, M11L and M11H) develop largely axisymmetric shell structures (Type 2), as conveyed by the representative Model 4 (with proper motion M11) of Fig. 2 (lower-right). This variation from Models 1-3 is attributed to the lessened capacity of the Galaxy to lock Fornax B into a more radial orbit and accretion by its host. These models have no tidal streams; in the case where tidal influence is most endorsed, with M11L and a lesser initial density of Fornax B (Model 5), small tidal plumes do develop (Fig. 4, lower-right panel).

\section{Discussion and Conclusions}
In this letter, we have demonstrated that the Fornax galaxy could be the remnant of a recent merger. We believe that Models 1 and 3 define two {\it bounding} cases, where a shell substructure, qualitatively similar to that observed in the Fornax dSph, is achieved for a merger occurring between $T_M$ = -3.5 and $T_M$ = -2.1 Gyr ago. Future photometry of these proposed shell regions may reveal their full distribution and thus allow further fine tuning of the merger time (and the initial density Fornax B); at this stage, however, the merger hypothesis put forward by Coleman et al. (2004) appears valid.
Furthermore, the morphology of this structure may be indicative of the orbital motion of Fornax; we anticipate the discovery of more substructure, but the current observations are most commensurate with the low-energy orbit of P07 (implying also D04, W08). We have additionally predicted tidal structure, similar to the well-studied stellar streams of the Sagittarius dSph (Marti\'nez-Delgado et al. 2004). Future observations of extant tails can provide constraints therefore on the proper motion of Fornax.
We suggest further that the maximal density regions of the substructure could represent the core of the companion, which has been not completely disrupted by the merger and subsequent relaxation. We intend that the models derived here feed a more detailed investigation, with emphasis on the gaseous evolution of the companion and a more accurate model for the potential field of Fornax.
\par
The stellar tails will likely be accompanied by an extensive dispersal of gas; the subsequent infall back to the core occurs over a timescale (Hernquist \& Spergel 1992, Hibbard \& Mihos 1995) that far exceeds the time since our postulated merger.
In spite of the significant disruption to the disk, the original stellar core of the companion survives in our simulations. While a rapid segregation of stars and gas may occur in an isolated galaxy (Weil \& Hernquist 1993), the tidal action on the remnant distribution may make gaseous infall less efficient, and thus the shell regions would have exhibited the merger-induced starburst rather than the galactic nuclei. 
Photometric analysis of stars within the inner and outer overdensities (Coleman et al. 2004, 2005) reveal a small age range centered on 2 Gyr; this truncation of star formation may be associated with the quick exhaustion of this cold gas under the action of tidal shocks, or instead a rapidly decreasing gas density (Kojima \& Noguchi 1997). Where infall does occur, the distinct velocity bimodality common to both the simulated Fornax B remnant and the metal-poor subpopulation of the Fornax dSph may support the concept of a young subpopulation recently introduced to the otherwise aged Fornax dSph.
\par
Our merger scenario may also add to the discussion on Fornax's globular clusters. Firstly, Buonanno et al. (1999) state a dichotomy in age between GCs with H4 being uniquely younger than clusters H1, H2, H3 and H5 by $\sim$3 Gyr. This contrasts with the opinion of Strader et al. (2003) where H5 is several years the youngest. In either case, the age discrepancy may be resolved by
the accretion of the younger GC {\it initially} in the companion galaxy. Based on observations of Forbes and Bridges (2011), several of the Galactic GCs have been traced to the Fornax-Leo-Sculptor system, and therefore may have recently been part of this companion; the tidal streams we have predicted may contain further GCs as evidence of this.
Secondly, the dynamical heating attributable to a merger 
event (Kazantzidis et al. 2008) could excite the orbits of the GCs.
Therefore it is possible that
the GCs have not since sunken to the core of the host 
and coalesced into its nucleus due to the past merger events of the Fornax dSph. Oh et al. (2000) calculated the timescale for mass accretion to the galaxy core (without tidal influences), as $\sim$1 Gyr, therefore implying that this heating mechanism may be required at several instances in the past. While this may appear presumptuous, it naturally supports the common observation that star formation in Fornax has occurred in a sequence of discontinuous starbursts. 

\acknowledgments
We gratefully acknowledge the suggestions of the anonymous referee that improved this manuscript.

\begin{figure}
	\epsscale{1.}
	\plotone{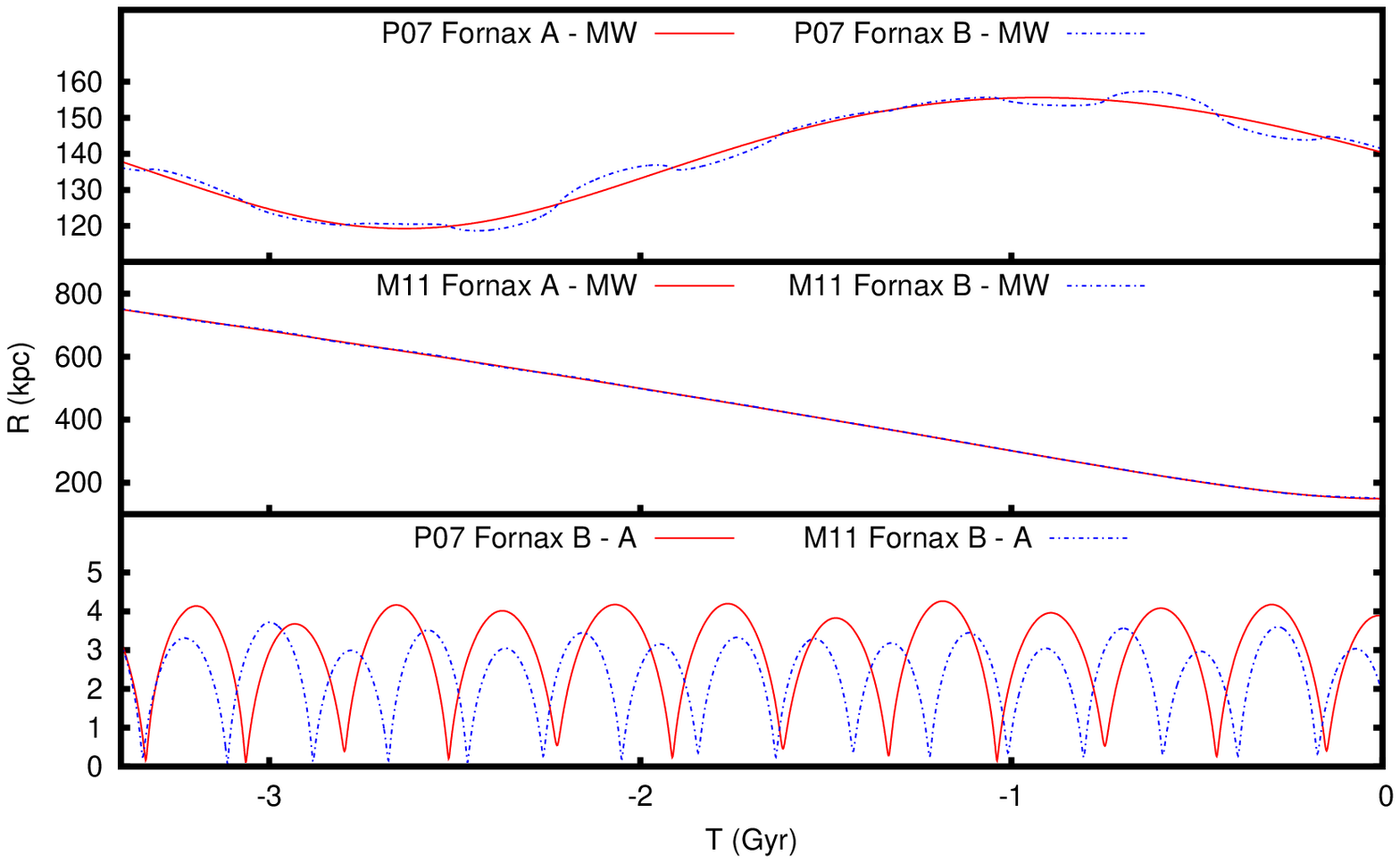}
	\caption{The radial distance between Fornax A and MW (red), Fornax B and MW (blue) for P07 (top panel) and M11 (middle panel); the lower panel shows the radial distance between Fornax A and B for P07 (Red) and M11 (blue).}
\end{figure}
\begin{figure}
	\epsscale{1.}
	\plotone{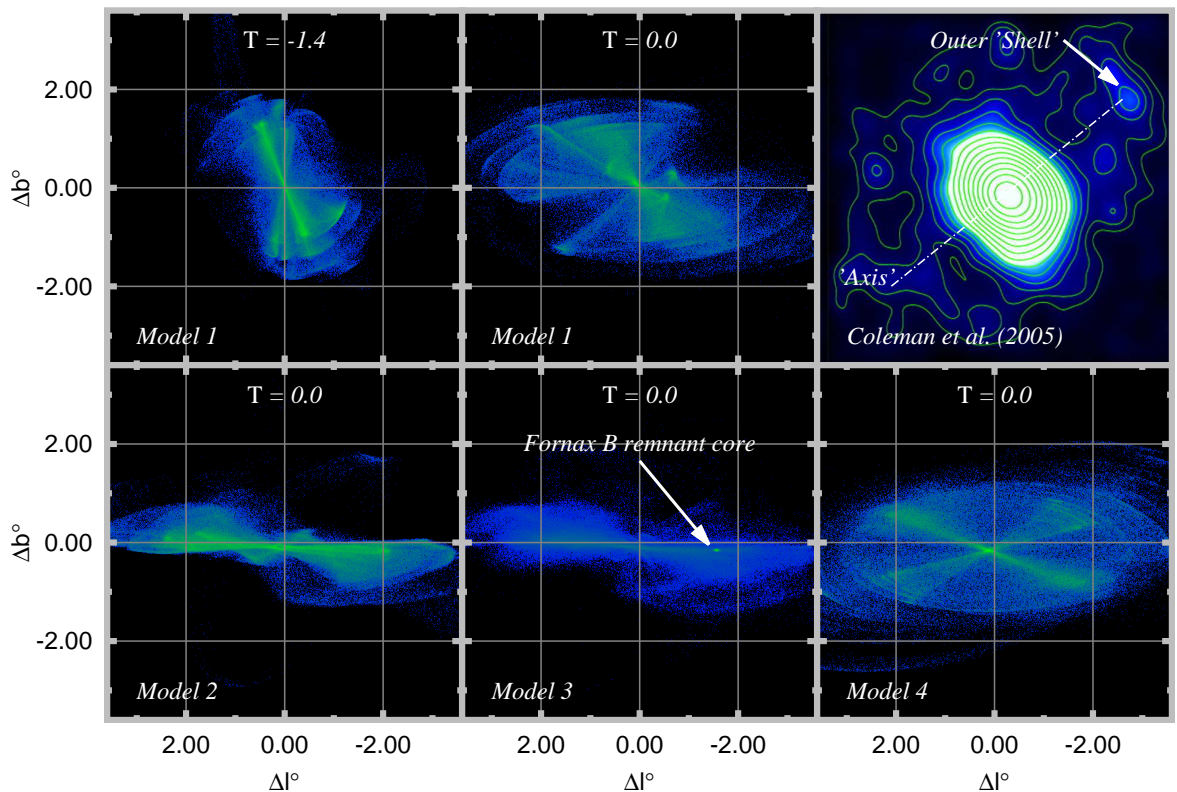}
	\caption{The stellar distribution of Fornax B as projected in galactic spherical coordinates and centered on the location of Fornax A. The time T (in units of Gyr) is shown at the top of each panel. The colour scale in each panel relates to the cumulative number of stars within each pixel of the 600x600 resolution image, for Model 1 (top-left, top-center), Model 2 (lower-left), Model 3 (lower-middle), and Model 4 (lower-right). For comparison, the Gaussian-convolved RGB stellar distribution of Fornax (Figure 23 of Coleman et al. 2005) is given in the top-right.}
\end{figure}

\begin{figure}
	\epsscale{1.}
	\plotone{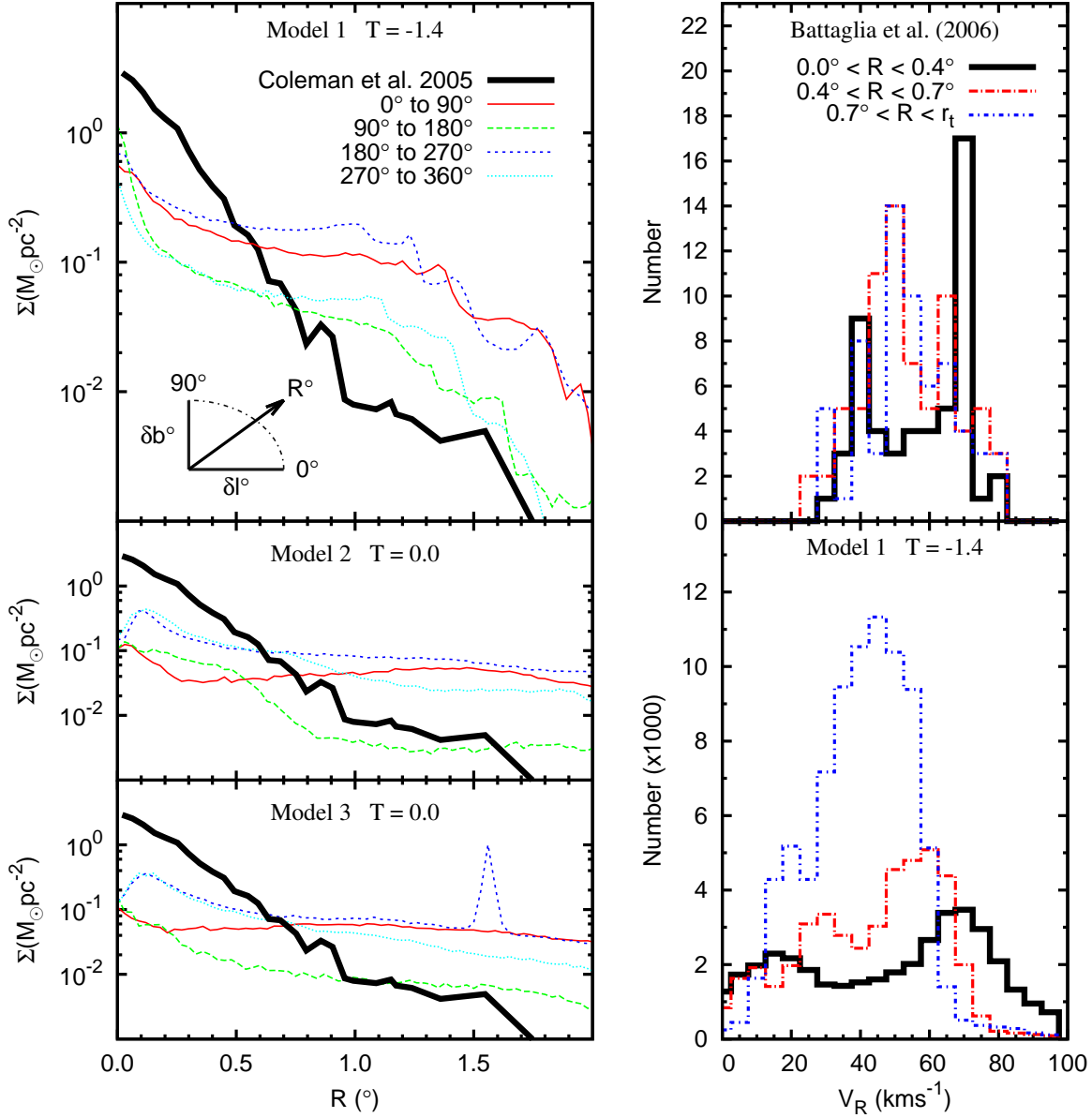}
	\caption{(Left) The stellar (mass) surface density of Fornax B for Models 1, 2 and 3, compared with stellar (number) density data (reproduced from Coleman et al. 2005). The distribution of stars from each N-body simulation is divided between four on-sky quadrants centered on Fornax A, at time T (in units of Gyr). (Right) The star count with Heliocentric Velocity for stars within distance bins up to the Fornax tidal radius, $r_t$ for (top) observational data reproduced from Battaglia et al. (2006), and (bottom) Model 1 from this work. }
\end{figure}

\begin{figure}
	\epsscale{1.}
	\plotone{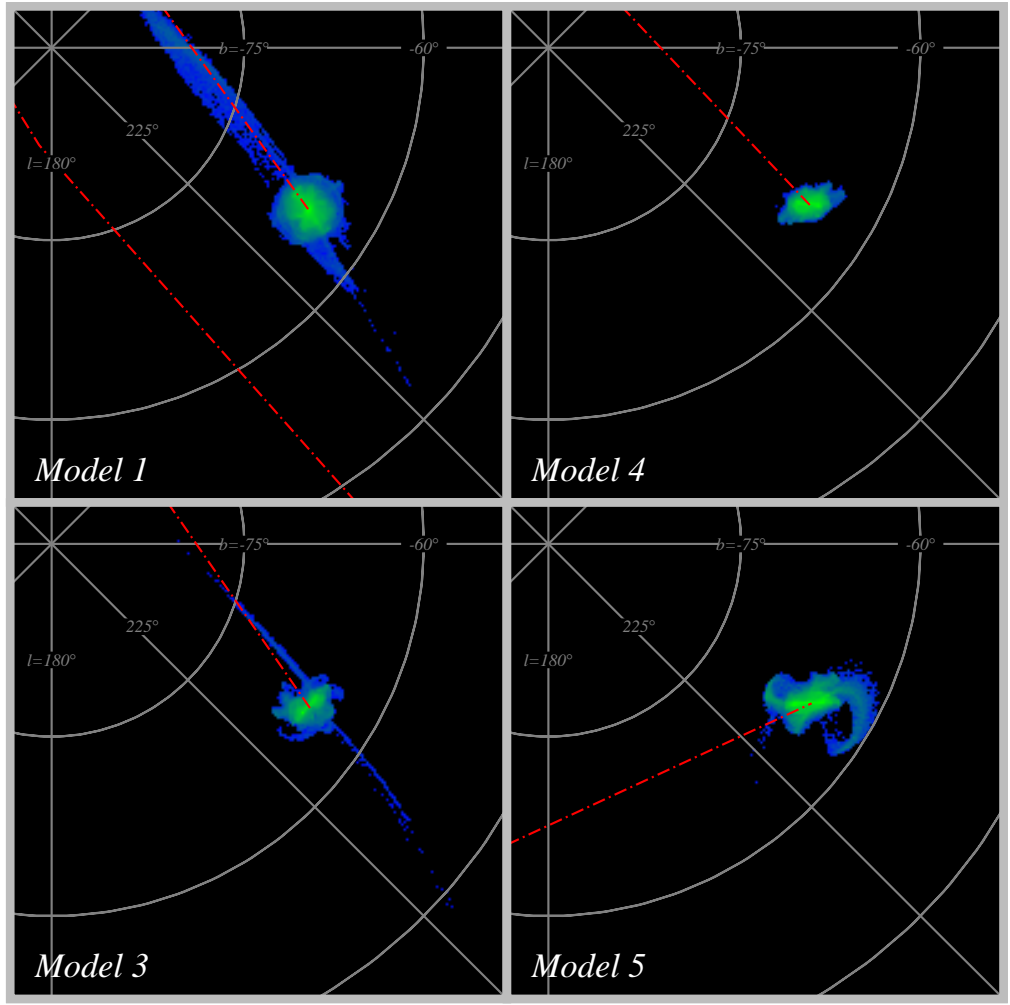}
	\caption{The stellar distribution of Fornax B at present time (T = 0) for Models 1, 3, 4 and 5. The colour scale is the same as used in Figure 2. The orbit of Fornax A is portrayed with the red dashed line.}
\end{figure}

\begin{deluxetable}{lcccccccc}
\tabletypesize{\scriptsize}
\tablecaption{Summary of model parameters\label{tbl-1}}
\tablewidth{0pt}
\tablehead{
\colhead{Model} & \colhead{Proper Motion} & \colhead{$T_M$} & \colhead{$V_X$} & \colhead{$V_Y$} & \colhead{$V_Z$} & \colhead{$r_{AB}$} & \colhead{$a_{\rm dm}$} & \colhead{$R_{\rm s, dw}$} \\
&  & \colhead{$Gyr$} & \colhead{$kms^{-1}$} & \colhead{$kms^{-1}$} & \colhead{$kms^{-1}$} & \colhead{$kpc$} & \colhead{$kpc$} & \colhead{$kpc$}
}
\startdata
1 & P07 & -3.5 & -137 & -80 & 103 & 3.75 & 0.645 & 0.6 \\
2 & P07 & -2.1 & -137 & -80 & 103 & 3.75 & 0.645 & 0.6 \\
3 & P07 & -2.1 & -137 & -80 & 103 & 3.75 & 0.43 & 0.4 \\
4 & M11 & -3.5 & -226 & -185 & 149 & 3.75 & 0.645 & 0.6 \\
5 & M11L & -3.5 & -125 & -96 & 99 & 3.75 & 0.645 & 0.6 \\
- & M11H & - & -326 & -274 & 199 & - & - & - \\

\enddata

\end{deluxetable}


\begin{thebibliography}{}

\bibitem[Battaglia et al. (2006)]{ba06} Battaglia, G. et al. 2006, A\&A, 459, 423 %
\bibitem[Bekki (2011)]{be11} Bekki, K. 2011, \mnras, 416, 2359 %
\bibitem[Bekki (2012)]{be12} Bekki, K. \& Yong, D. 2012, \mnras, 419, 2063 %
\bibitem[Binney \& Tremaine (2008)]{bi08} Binney, J. \& Tremaine, S. 2008, Galactic Dynamics, Princeton Univ. Press (Princeton, NJ) %
\bibitem[Buonanno et al. (1999)]{bu99} Buonanno, R. et al., 1999, \aj, 118, 1671 %
\bibitem[Coleman et al. (2004)]{co04} Coleman, M. G. et al., 2004, \aj, 127, 832 %
\bibitem[Coleman et al. (2005)]{co05} Coleman, M. G. et al. 2005, \aj, 129, 1443 %
\bibitem[Cowsik et al. (2009)]{co09} Cowsik, R. et al., 2009, \apj, 699, 1389 %
\bibitem[Dinescu et al. (2004)]{di04} Dinescu, D. I. et al., 2004, \aj, 128, 687 %
\bibitem[Diaz \& Bekki (2012)]{di12} Diaz, J. \& Bekki, K. 2012, \mnras, 413, 2015 %
\bibitem[Forbes \& Bridges (2010)]{fo10} Forbes, D. A. \& Bridges, B., 2011, \mnras, 404, 1203 %
\bibitem[Goerdt et al. (2006)]{go06} Goerdt, T. et al., 2006, \mnras, 368, 1073 %
\bibitem[Hernquist \& Quinn (1989)]{he89} Hernquist, L. \& Quinn, P., 1989, \apj, 342, 1 %
\bibitem[Hernquist (1990)]{he90} Hernquist, L., 1990, \apj, 356, 359 %
\bibitem[Hernquist (1992)]{he92} Hernquist, L. \& Spergel, D. N. 1992, \apj, 399, L117  %
\bibitem[Hibbard \& Mihos (1995)]{hi95} Hibbard, J. E. \& Mihos, J. C. 1995, \aj, 110, 140 %
\bibitem[Irwin \& Hatzidimitriou (1995)]{ir95} Irwin, M., \& Hatzidimitriou, D. 1995, \mnras, 277, 1354 %
\bibitem[Kazantzidis et al. (2008)]{ka08} Kazantzidis, S et al., 2008, \apj, 688, 254 %
\bibitem[Kirby et al. (2011)]{ki11} Kirby, E. N. et al., 2011, \apj, 727, 79 %
\bibitem[Kojima \& Noguchi (1997)]{ko97} Kojima, M. \& Noguchi, M. 1997, \apj, 481, 132 %
\bibitem[Marti\'nez-Delgado et al. (2004)]{ma04} Marti\'nez-Delgado, D. et al., 2004, \apj, 601, 242 %
\bibitem[Mateo et al. (1991)]{ma91} Mateo, M. et al., 1991, \aj, 102, 914 %
\bibitem[M\'endez et al. (2011)]{me11} M\'endez, R. A., Costa, E., Gallart, C. et al., 2011, \aj, 142, 93 %
\bibitem[Miyamoto \& Nagai (1975)]{mi75} Miyamoto, M. \& Nagai, R. 1975, \pasj, 27, 533 %
\bibitem[Murai \& Fuijimoto (1980)]{mu80} Murai, T., \& Fujimoto, M. 1980, \pasj, 32, 581 %
\bibitem[Navarro et al. (1996)]{na96} Navarro, J., F. et al., 1996, \apj, 490, 493 %
\bibitem[Oh et al. (2000)]{oh00} Oh, K. S. et al., 2000, \aj, 531, 727 %
\bibitem[Olszewski et al. (2006)]{ol06} Olszewski, E. W., Mateo, M., \& Harris, J., et al., 2006, \aj, 131, 912 %
\bibitem[Piatek et al. (2007)]{pi07} Piatek, S. et al., 2007, \aj, 133, 818 %
\bibitem[Prieur (1990)]{pr90} Prieur, J. -L. 1990, in Dynamic and Interactions of Galaxies, ed. R. Wielen (Springer-Verlag, Berlin) %
\bibitem[Quinn (1984)]{qu84} Quinn, P. 1984, \apj, 279, 596 %
\bibitem[Salucci \& Burkert (2000)]{sb00} Salucci, P. \& Burkert, A. 2000, \apj, 312, L27 %
\bibitem[Savianne et al. (2000)]{sa00} Saviane, I. et al., 2000, A\&A, 355, 56 %
\bibitem[Stetson et al. (1998)]{st98} Stetson, P.B. et al. 1998, \pasp, 110, 533 %
\bibitem[Strader et al. (2003)]{st03} Strader, J. et al., 2003, \aj, 125, 1291 %
\bibitem[Tsujimoto (2011)]{ts11} Tsujimoto, T. 2011, \apj, 736, 113 %
\bibitem[van den Bergh (1998)]{va98} van den Bergh, A. 1998, \apj, 492, 41 %
\bibitem[van den Bergh (1999)]{va99} van den Bergh, S, 1999, in IAU Symp. 192, The Stellar Content of Local Group Galaxies, ed. P. Whitelock \& R. Cannon (Cambridge: Cambridge Univ. Press), 3  %
\bibitem[Walker et al. (2007)]{wa07} Walker, M. G. et al., 2007, \apj, 667, L53 %
\bibitem[Walker et al. (2008)]{wa08} Walker, M. G. et al., 2008, \apj, 688, L75 %
\bibitem[Weil \& Hernquist (1993)]{we93} Weil, M. L. \& Hernquist, L., 1993, \apj, 405, 142 %
\bibitem[Weil et al. (1996)]{we96} Weil, M. L. et al., 1996, \apj, 490, 664 %

\end{thebibliography}
\end{document}